\documentclass[12pt,psfig]{article}
\usepackage{amssymb}
\usepackage{amsmath,amsthm}
\usepackage{amsfonts}
\usepackage{graphicx}
\usepackage{braket}
\usepackage{graphics}
\usepackage{indentfirst}
\usepackage{hyperref}

\begin{document}
%\title{ui}\maketitle
\begin{center}\Large\textbf{Open String Pair Production 
from the Interacting Dressed-Angled D-Strings }
\end{center}
\vspace{0.75cm}
\begin{center}
\large{Hamidreza Daniali and \large Davoud Kamani}
\end{center}
\begin{center}
\textsl{\small{Department of Physics, 
Amirkabir University of
Technology (Tehran Polytechnic), Iran \\
P.O.Box: 15875-4413 \\
e-mails: hrdl@aut.ac.ir , kamani@aut.ac.ir \\}}
\end{center}
\vspace{0.5cm}

\begin{abstract}

We calculate the rate of the open string pair 
production from the interaction of two 
non-intersecting D-strings at 
angle. The D-strings have been 
equipped with the $U(1)$ gauge 
potentials in the presence of the Kalb-Ramond field. 
This rate will be discussed when the D-strings 
intersect each other. Our computations is in 
the framework of the bosonic string theory.

\end{abstract}



\newpage

%%%%%%%%%%%%%%%%%%%%%%%%%%%%%%%%
\section{Introduction}
\label{100}

The D-branes are essential objects for 
various descriptions of the string theory  
\cite{1, 2, 3}. 
For computing the lowest order stringy 
interaction amplitude of the D-branes 
in the closed string channel, 
one can apply the boundary state 
formalism \cite{4}-\cite{17}. 
Equivalently, it is possible to calculate 
the same amplitude via the one-loop annulus
of open string \cite{1, 14, 18}.
By adding dynamics, background
fields and different internal gauge 
potentials to the D-branes, 
one can reveal more properties of them 
\cite{11}-\cite{15}, \cite{19}-\cite{23}. 
For the branes with a large separation, the interaction
occurs via the exchange of the massless closed strings, 
which is attractive \cite{14, 20, 23}. 
However, in the small distance limit, 
the nature of the branes interaction 
(attraction or repulsion) completely is ambiguous. 
As we shall see, the open string channel provides 
the most accurate description for such configurations.

The electric field on a D-brane 
provides the necessary energy for 
creating the open string pairs, 
analogous to the Schwinger pair production \cite{24}. 
Thus, in the presence of the electric fields, 
the interaction amplitude of the D-branes
possesses an imaginary part, which leads to 
the open string pair production.
In fact, the production rate of the 
open string pairs is infinitesimally small. However, 
when the magnetic fields are applied to
the system of the branes this rate is extremely modified
\cite{25}-\cite{29}.

The open string pair production, similar to the 
other investigations of the branes, has been usually
studied for the parallel branes. However,
the branes at angles have been also applied 
in various subjects, e.g. in the super
Yang-Mills theories \cite{30, 31},
in the modeling of the black holes \cite{32}, 
in the non-chiral supersymmetric field theories 
\cite{33}-\cite{35}, and so on.
Therefore, we shall study the open string pair production 
via two angled-dressed D-strings (D1-branes)
which are not intersecting.
The D-strings have been dressed with 
the different $U(1)$ gauge 
potentials and live in a nonzero Kalb-Ramond 
background field. The production 
rate will be also investigated for two intersecting D-strings.
Our calculations is 
in the context of the bosonic string theory

This paper is organized as follows. 
In Sec. \ref{200}, we obtain the boundary state, 
associated with a skew-dressed D-string
in the presence of the Kalb-Ramond field. 
In Sec. \ref{300}, at first, we calculate the interaction 
amplitude of a system of two D-strings at angle 
with a nonzero distance. Then, we compute 
the pair production rate of the open strings
from the interaction of this system. 
In the subsection \ref{301}, we investigate this 
rate for the intersecting D-strings. 
The conclusions will be given in Sec. \ref{400}.

%%%%%%%%%%%%%%%%%%%%%%%%%%%%%%%%
\section{The boundary state 
corresponding to an oblique-dressed D-string}
\label{200}
 
In order to compute the open string pair production, 
it is necessary to calculate the interaction amplitude. 
The angled D-strings (D1-branes) of 
the setup have been dressed with the 
$U(1)$ gauge potentials in the 
presence of a constant background $B$-field. 
The interaction will be computed 
in the closed string channel. Thus, we  
use the boundary state formalism. 
Hence, at first we construct the boundary state, 
corresponding to a D-string along the 
$x_1$-direction. So, we begin with the 
following string action
\begin{eqnarray}
\label{1}
S &=& -\dfrac{1}{4 \pi \alpha^\prime} \int_\Sigma 
{\rm d}^2 \sigma \left(\sqrt{-h} h^{ab} G_{\mu\nu}(X) 
+ \epsilon^{ab} B_{\mu\nu}(X) \right)
\partial_a X^\mu \partial_b X^\nu 
\nonumber\\
&+&\dfrac{1}{2\pi \alpha^\prime} \int_{\partial\Sigma} 
{\rm d}\sigma A_\alpha \partial_\sigma X^\alpha,
\end{eqnarray}
where $\mu,\nu$ are the spacetime indices and $h^{ab}$, 
with $a,b\in \{\tau, \sigma\}$, represents the metric of the 
closed string worldsheet. The 
indices $\alpha , \beta \in \{0,1\}$ 
indicate the worldsheet of the D-string.
For the next purposes the indices 
$i , j \in \{2, \cdots, d-1\}$ 
will represent transverse directions to the D-string.
We shall apply the flat metric  
$G_{\mu\nu} = \eta_{\mu\nu} = {\rm diag} (-1 , 1, \cdots ,1 )$ 
for the spacetime, a constant Kalb-Ramond field 
$B_{\mu\nu}$, and the Landau gauge 
$A_\alpha = -2^{-1} F_{\alpha\beta} X^\beta$ 
with the constant field strength $F_{\alpha\beta}$. 
In fact, this gauge dedicates a squared structure 
to the action \eqref{1}.

Setting the action variation to zero 
yields the equation of motion
and also the following boundary state equations
\begin{eqnarray} \label{2}
&&\left[\partial_\tau X^0(\sigma,\tau) - \mathcal{E} 
\partial_\sigma 
X^1(\sigma,\tau) \right]_{\tau=0} | B'\rangle 
= 0 ,\nonumber\\
&&\left[\partial_\tau X^1(\sigma,\tau) - \mathcal{E} 
\partial_\sigma 
X^0 (\sigma,\tau)\right]_{\tau=0} | B'\rangle 
= 0 ,\nonumber\\
&&\left[X^i(\sigma,\tau) - y^i\right]_{\tau=0} | B'\rangle 
= 0,
\label{2}
\end{eqnarray}
where  $\mathcal{E} \equiv F_{01} - B_{01}$, the parameters 
$y^i$ indicate the position of the D-string,
and $| B'\rangle$ represents the 
corresponding boundary state of it.

Now consider a D-string parallel to the $x^1x^2$-plane, 
which makes the angle $\phi$ with the $x^1$-direction. 
By utilizing the well-known closed string mode expansion, 
the zero-mode part of the boundary state equations 
for the oblique D-string become
\begin{eqnarray}
\label{3}
&&\hat{p}^0 |B\rangle_0 = 0, 
\nonumber\\
&&\left(\hat{p}^1 \cos \phi + \hat{p}^2 \sin 
\phi\right)|B\rangle_0 = 0,
\nonumber\\
&&\left[ (\hat{x}^2 - y^2) \cos \phi -  
(\hat{x}^1 - y^1) \sin \phi\right] 
|B\rangle_0 = 0, 
\nonumber\\
&&(\hat{x}^{i'} - y^{i'}) |B\rangle_0 = 0,
\end{eqnarray}
in which $i'\in \{3,4, \cdots, d-1\}$, and 
$| B\rangle$ shows the boundary state
of the oblique D-string. We employed the decomposition 
$| B\rangle = | B\rangle_0 \otimes | B\rangle_{\rm osc}$. 
Using the quantum mechanical technics, the zero-mode 
portion of the boundary state finds the following solution
\begin{eqnarray}
\label{4}
|B\rangle_0 &=& \frac{T_1}{2}\delta
\left[ (\hat{x}^2 - y^2) \cos \phi 
-  (\hat{x}^1 - y^1) \sin \phi\right] |p^0 
= 0\rangle|p^1 = 0\rangle |p^2 = 0\rangle 
\nonumber\\
&\times&\left( \prod_{i ' =3}^{d-1} \delta 
(\hat{x}^{i'} - y^{i'}) |p^{i'} = 0\rangle\right),
\end{eqnarray}
where $T_1=g^{-1}_s\sqrt{\pi}\;2^{(10-d)/4}
(4\pi^2 \alpha')^{(d-6)/4}$ 
is the tension of the D-string.

The oscillating parts of the boundary state equations
can be written in a unified form, i.e., 
$\left[\alpha^\mu_n + \Omega^\mu_{\ \nu} (\phi, \mathcal{E}) 
\tilde{\alpha}^\nu_{-n}\right] |B\rangle_{\rm osc} = 0$, 
where $\Omega^\mu_{\ \nu} = 
\Big( \mathcal{M}^{\alpha'}_{\ \beta'}(\phi, \mathcal{E}), 
- \delta^{i'}_{ \ j'} \Big)$ 
with $\alpha',\beta'\in \{0,1,2\}$, and
\begin{eqnarray}
\label{5}
\mathcal{M}^{\alpha'}_{\ \beta'}(\phi, \mathcal{E}) 
&\equiv & \frac{1}{1 - \mathcal{E}^2} 
\begin{bmatrix}
1+\mathcal{E}^2       & -2 \mathcal{E} \cos\phi 
& -2 \mathcal{E} \sin\phi \\
-2 \mathcal{E} \cos\phi       
& \mathcal{E}^2 + \cos(2\phi) & \sin(2\phi)\\
-2 \mathcal{E} \sin\phi      
& \sin(2\phi) & \mathcal{E}^2 - \cos(2\phi)
\end{bmatrix}.
\end{eqnarray}
By employing the coherent state method, we acquire the 
solution 
\begin{eqnarray}
|B\rangle_{\rm osc} = \sqrt{1-\mathcal{E}^2} 
\exp \left( - \sum_{n=1}^{\infty}\frac{1}{n}
\alpha^\mu_{-n} \Omega_{\mu \nu} 
\tilde{\alpha}^{\nu}_{-n}\right)|0\rangle 
\otimes |\tilde{0}\rangle.
\nonumber
\end{eqnarray}
The prefactor $\sqrt{1-\mathcal{E}^2}$ comes 
from disk partition function \cite{9}.
For the future purposes, we should note that 
the matrix $\mathcal{M}$ is orthogonal 
and hence $\Omega$ also is an orthogonal matrix. 

In the bosonic string theory, the direct product 
$|B\rangle_{\rm tot}= |B\rangle_{\rm osc} 
\otimes |B\rangle_{0} \otimes |B\rangle_{\rm g}$ 
exhibits the total boundary state, associated with the 
D-string, where $|B\rangle_{\rm g}$ 
is the well-known boundary state of the conformal ghosts. 

%%%%%%%%%%%%%%%%%%%%%%%%%%%%%%%%%%%%
\section{Open string pair production from 
the non-intersecting-angled D-strings}
\label{300}

In this section, we extract the 
rate of the open string pair 
creation via the interaction amplitude 
of two D-strings at angles. 
For preserving the generality, assume 
that the electric fields and angles of the   
D-strings with the $x^1$-direction are different. 
Thus, the subscripts $(1)$ and $(2)$ will be used to 
show these differences.

In the closed string channel, the
interaction of two D-branes
takes place by exchanging closed
strings between the branes.
The geometry of the worldsheet of the exchanged 
closed string is a cylinder with $\tau$ as the
coordinate along the length of the
cylinder, $ 0\le \tau \le t$, and $\sigma$ as the periodic
coordinate,  i.e. $ 0 \le \sigma\le \pi$. 
The interaction amplitude can be computed via
the overlap of the two boundary states, through the closed
string propagator 
$\mathcal{A}_{\rm closed} = 
\pi \alpha' \int_0^\infty {\rm d}t\ \  _{\rm tot}\langle 
B_{(1)}|\exp(-t\mathcal{H}) |B_{(2)}\rangle_{\rm tot}$, 
where $\mathcal{H}$ is the total Hamiltonian of the 
closed string. It includes the ghost and matter parts.
After long calculations one finds
\begin{eqnarray}
\label{6}
\mathcal{A}_{\rm closed} 
&=&  V_1 \frac{
\;\sqrt{\left(1-\mathcal{E}_{(1)}^2\right)
\left(1-\mathcal{E}_{(2)}^2\right)}}{\sqrt{8\pi^2\alpha'}g_s^2 
|\sin\Phi|} \int_0^\infty {\rm d}t \ 
\frac{e^{(d-2)\pi t/12}}{t^{(d-3)/2}}
\exp \left( -\frac{\mathbf{Y}^2}{2 \pi\alpha't} \right)
\nonumber \\
&\times& \prod_{n=1}^\infty\left[ 
\frac{(1-e^{-2n\pi t})^{4-d}}{1-2e^{-2n\pi t} \left( 
\frac{2\left(\cos\Phi -\mathcal{E}_{(1)}\mathcal{E}_{(2)} 
\right)^2}{\left(1-\mathcal{E}_{(1)}^2\right)
\left(1-\mathcal{E}_{(2)}^2\right)}-1\right) 
+ e^{-4n\pi t}}\right],
\end{eqnarray} 
with $\Phi \equiv \phi_{(2)}-\phi_{(1)}$ 
is the angle between the D-strings, and 
$\mathbf{Y}^2 \equiv \sum_{j'=3}^{d-1} 
\left( y_1^{j'}-y_2^{j'}\right)^2$ is the 
square distance between them. 

Let us analyze the amplitude \eqref{6}.
The factor $e^{(d-2)\pi t/12}$ comes from  
the Hamiltonian regularization, 
and $V_1 = 2 \pi \delta(0)$ is the 
length of the D-string. The advent of 
$|\sin \Phi|$ in the denominator, 
originates from the property 
$\delta(ax) = |a|^{-1}\delta(x)$. 
This elucidates that the amplitude  
\eqref{6} is valid only for 
$\Phi \ne 0$, and the parallel 
case is different and cannot be 
obtained from this by setting $\Phi =0$.  
In the latter configuration, for example, the $\mathbf Y$ 
is $(d - 2)$-dimensional and $V_1$ should be 
substituted by $V_2 = (2\pi)^2 \delta^2(0)$.
The position-dependent factor clarifies that the 
interaction amplitude is exponentially 
damped by increasing the distance between 
the D-strings. 
The infinite product is the contributions of the oscillatory 
part and the conformal ghosts portion. Precisely,
the contribution of the conformal
ghosts is $(1-e^{-2n\pi t})^{2}$, 
and the contribution of the 
transverse directions to both D-strings  
is $(1-e^{-2n\pi t})^{3-d}$. 
The expression in the denominator of the infinite 
product can be rewritten in the form  
$(1-\gamma e^{-2n\pi t})(1-\gamma' e^{-2n \pi t})$,
where $\gamma \gamma' =1$. Therefore, we receive  
\begin{eqnarray}
\label{7}
\frac{1}{2}\left(\gamma + \gamma'^{-1}\right)  
= \frac{2\left(\cos\Phi-\mathcal{E}_{(1)}\mathcal{E}_{(2)} 
\right)^2}{\left(1-\mathcal{E}_{(1)}^2\right)
\left(1-\mathcal{E}_{(2)}^2\right)} -1.
\end{eqnarray} 

By writing $\gamma \equiv e^{2i\pi \nu}$,
Eq. \eqref{7} takes the feature 
\begin{eqnarray}
\label{8}
\cos\pi\nu = \frac{\cos\Phi 
-\mathcal{E}_{(1)}\mathcal{E}_{(2)} 
}{\sqrt{\left(1-\mathcal{E}_{(1)}^2\right)
\left(1-\mathcal{E}_{(2)}^2\right)}}.
\end{eqnarray} 
In the case of parallel D-strings 
there is $\cos \Phi =1$. 
For this configuration we have 
$1 -\mathcal{E}_{(1)}\mathcal{E}_{(2)}> 
\sqrt{\left(1-\mathcal{E}_{(1)}^2\right)
\left(1-\mathcal{E}_{(2)}^2\right)}$,
e.g. see \cite{25}-\cite{29}.
This inequality elaborates that $\nu$ is 
a pure imaginary quantity. 
This is an essential element for 
calculating the production rate of the open string pairs.
Hence, we utilize  
$\nu= i\tilde\nu$ in which $0< \tilde\nu<\infty$. 
Consequently, Eq. \eqref{8} takes the feature 
\begin{eqnarray}
\cos\Phi = \mathcal{E}_{(1)}\mathcal{E}_{(2)}
+\cosh \pi {\tilde \nu}\;
\sqrt{\left(1-\mathcal{E}_{(1)}^2\right)
\left(1-\mathcal{E}_{(2)}^2\right)}. 
\label{9}
\end{eqnarray}
We observe that the perpendicularity of 
the two D-strings is not possible unless for the case 
$\mathcal{E}_{(1)}\mathcal{E}_{(2)}<0$. 
Since $\cosh\pi{\tilde \nu} \geq 1$, we obtain  
$\mathcal{E}_{(1)}^2+\mathcal{E}_{(2)}^2 \geq 2  
\mathcal{E}_{(1)}\mathcal{E}_{(2)} \cos \Phi + \sin^2 \Phi$.
This condition imposes some restrictions on the angle $\Phi$
and the electric fields. Precisely, for the 
given electric fields $\mathcal{E}_{(1)}$
and $\mathcal{E}_{(2)}$, for occurring the 
open string pair production 
the possible range of the angle $\Phi$ is 
defined by  
$\cos \Phi > x_+$ or $\cos \Phi < x_-$, in which  
$x_\pm \equiv \mathcal{E}_{(1)}\mathcal{E}_{(2)} \pm \
\sqrt{\left(1-\mathcal{E}_{(1)}^2\right)
\left(1-\mathcal{E}_{(2)}^2\right)}$.
In fact, the electric field of a D-string is parallel to 
that D-string. Thus, the angle $\Phi$ represents 
the angle between the electric fields too.
Hence, for happening the pair creation of open strings,
the electric fields should possess special 
alignments with each other.

By utilizing the Jacobi $\Theta$- 
and Dedekind $\eta$-functions, 
the amplitude finds the following feature
\begin{eqnarray}
\label{10}
\mathcal A_{\rm closed} &=& iV_1 
 \frac{\sinh \pi{\tilde \nu}\sqrt{\left(1-\mathcal{E}_{(1)}^2\right)
\left(1-\mathcal{E}_{(2)}^2\right)}}
{\sqrt{2\pi^2\alpha' }g_s^2|\sin \Phi|} \int_0^\infty 
\frac{{\rm d}t}{t^{(d-3)/2}} 
\frac{\exp\left(-\frac{\mathbf{Y}^2}{2\pi\alpha' t}\right) }
{\eta^{d-5}(it)
\Theta_1(i\tilde{\nu} | it)}  
\end{eqnarray}
This amplitude is real.
For the large D-strings separation, which is 
equivalent to $t \rightarrow \infty$,
the dominant contribution to the 
interaction arises from the massless closed strings.
In this case we receive 
$\mathcal A_{\rm closed}|_{t\rightarrow\infty}>0$, 
which represents an attractive force.

We now consider the small ``$t$''-integration. 
In this limit, the sign of the denominator in 
the infinite product of Eq. \eqref{6} becomes 
negative. Hence, the sign of the infinite product
is unclear. This implies that 
any statement regarding the nature of the 
interaction (attraction or repulsion)
is ambiguous. In fact, for the 
infinitesimal value of ``$t$'',
the open string amplitude is more appropriate.
Thus, let us employ the Jacobi transformation 
$t \rightarrow  1/t$ to convert the closed string 
amplitude \eqref{10} to the open string annulus amplitude. 
With the use of
\begin{eqnarray}
\Theta_1(\hat\nu|\hat\tau) 
= i \frac{e^{-i\pi \hat\nu^2/\hat\tau}}
{\sqrt{-i\hat\tau}} \ \Theta_1 
\left( \frac{\hat\nu}{\hat\tau}\right|\left. 
- \frac{1}{\hat\tau}\right), \qquad\eta(\hat\tau)
= \frac{1}{\sqrt{-i\hat\tau}}\  
\eta\left(-\frac{1}{\hat\tau}\right),
\end{eqnarray}
we obtain  
\begin{eqnarray}
\mathcal A_{\rm open} 
&=& V_1\frac{\sqrt{\left(1-\mathcal{E}_{(1)}^2\right)
\left(1-\mathcal{E}_{(2)}^2\right)}}{\sqrt{2\pi^2\alpha' }
g_s^2|\sin \Phi|} 
\sinh \pi{\tilde \nu} \int_0^\infty \frac{{\rm d}t}{t^{3/2}} 
\frac{\exp\left(-\frac{\mathbf{Y}^2t}{2\pi\alpha' }\right) 
e^{-\pi \tilde{\nu}^2t} }{\eta^{d-5}(it)
\Theta_1(\tilde{\nu}t | it)} 
 \label{12}
\end{eqnarray}	

For the next purposes, we write the amplitude 
\eqref{12} in the form 
\begin{eqnarray}
\label{13}
\mathcal A_{\rm open} &=& 
V_1\frac{\sqrt{\left(1-\mathcal{E}_{(1)}^2\right)
\left(1-\mathcal{E}_{(2)}^2\right)}}
{\sqrt{8\pi^2\alpha' }g_s^2|\sin \Phi|} \sinh \pi{\tilde \nu} \int_0^\infty 
{\rm d}t \frac{e^{-2\mathcal M^2 t}}{t^{3/2}
\sin(\tilde \nu \pi t)}\nonumber \\
&\times& \prod_{n=1}^\infty \frac{(1-e^{-2n\pi t})^{4-d}}{1-
2  \cos(2\pi \tilde{\nu} t)e^{-2n\pi t}  + e^{-4n \pi t}} 
\end{eqnarray}
where the effective mass of the open string,
stretched between the D-strings, is given by 
\begin{eqnarray}
\label{14}
\mathcal{M}^2 \equiv 
\frac{1}{4\pi \alpha'} \left\{\mathbf{Y}^2 
+ \pi^2 \alpha' \left[2\tilde\nu^2 - \frac{d-2}{6}\right] \right\}.
\end{eqnarray} 
When $\mathbf{Y} < \pi\sqrt{\alpha' \left[\frac{d-2}{6} 
-2\tilde\nu^2 \right]}$ we receive the tachyonic shift. 
In this case, the integrand of  
Eq. \eqref{13} for $t \rightarrow \infty$ diverges,
which indicates the tachyonic instability. 
Therefore, a phase transition will take place through 
the tachyon condensation \cite{36}.
We discard the tachyonic shift, i.e. 
by choosing appropriate values for the 
setup parameters, we apply  
$\mathcal{M}^2\geq 0$. Hence, the formulation of 
the open string pair production is permitted.

In the open string channel, the  
small D-strings separation is equivalent 
to the limit $t \to \infty$. In this case, 
all factors in the integrand 
of Eq. \eqref{13} are positive, while 
the value of the factor 
$\sin (\tilde\nu\pi t)$ belongs 
to the interval $(-1,1)$. 
This factor leads to an infinite number 
of simple poles along the 
positive $t$-axis, $t_m = \frac{m}{\tilde\nu}$ 
where $m= 1,2,3,\cdots$. Each pole separately
exhibits the creation of a pair of open strings 
and also the decay of the system. 
Besides, the amplitude includes an imaginary part. 
By calculating the residue,
following Refs. \cite{37, 38}, and 
using the well-known Schwinger formula 
$\mathcal{W} = - 2 V_{1+1}^{-1}\ {\rm Im} \ 
\mathcal A_{\rm open}$, 
we find the pair production rate per unit 
worldsheet area $V_{1+1}$ as in the following 
\begin{eqnarray}
\label{15}
\mathcal{W} 
&=& \frac{V_1}{V_{1+1}} \frac{\sqrt{\left(
1-\mathcal{E}_{(1)}^2\right)
\left(1-\mathcal{E}_{(2)}^2\right)}}
{\sqrt{2\pi^2\alpha' }g_s^2|\sin \Phi|} 
{\tilde\nu}^{1/2}\sinh \pi{\tilde \nu} \; 
\sum_{m=1}^\infty \bigg{\{}
\frac{(-1)^{m+1}}{m^{3/2}}
\nonumber \\
&\times& \exp\left(-\frac{2 m }
{\tilde\nu}\mathcal{M}^2\right) 
\prod_{n=1}^\infty \left[ 1- \exp\left(-\frac{2n\pi m}
{\tilde\nu}\right)\right]^{2-d}\bigg{\}}.
\end{eqnarray}
For the finite distance of the D-strings, especially 
when they are near to each other, 
this quantity elaborates the creation 
of the open string pairs and
consequently the decay of the system.
Note that the open strings production 
between the D-branes has resemblances with the
Casimir effect.

Eq. \eqref{9} implies that the decay rate \eqref{15} is 
a complicated function of the angle $\Phi$ and 
the electric fields on the D-strings.
We observe that by increasing the D-strings distance
$\mathbf{Y}$, the pair production rate decreases. 
In other words, for the large value of 
the distance $\mathbf{Y}$, 
the mass of each pair obviously becomes large. 
Consequently,
the probability of producing the heavy open 
string pairs is small. 
Besides, for the case $d \geq 3$, as it ought to be 
for non-intersecting-angled D-strings, by increasing the 
${\tilde \nu}$ the production rate also increases.

According to Eq. \eqref{9}, the electric 
fields cannot simultaneously vanish.
Therefore, in the decay rate \eqref{15}
at least one of the electric fluxes
should be nonzero. From the physical point 
of view, at least an electric flux
is prominently needed to polarize 
the region between the D-strings. 
Otherwise, the pair creation of the 
open strings does not occur.

For the case $\mathcal{E}_{(1)} \rightarrow 1$ 
or $\mathcal{E}_{(2)} \rightarrow 1$ or 
both, the production rate begins to diverge, 
and the pair production instability occurs. 
In the limit $\tilde\nu \rightarrow 0$, which 
takes place for a setup with the following relation 
among its parameters   
$\cos \Phi \approx \mathcal{E}_{(1)}\mathcal{E}_{(2)} 
+ \sqrt{(1-\mathcal{E}^2_{(1)})(1-\mathcal{E}^2_{(2)})}$, 
we acquire the minimum rate 
\begin{eqnarray}
\mathcal{W}_{\tilde\nu \rightarrow 0} \approx 
 \frac{V_1}{V_{1+1}} \frac{\sqrt{\left(1-\mathcal{E}_{(1)}^2\right)
\left(1-\mathcal{E}_{(2)}^2\right)}}
{\sqrt{2\pi^2\alpha' }g_s^2|\sin \Phi|} \tilde\nu^{3/2}
\exp\left(-\frac{2}{\tilde\nu}\mathcal{M}^2\right).
\end{eqnarray}
This rate prominently is infinitesimal and does not possess 
any physical consequence.

%%%%%%%%%%%%%%%%%%%%%%%%%%%%%%%%%%%%%%%%%%%%%%%%%
\subsection{Pair production rate of photons  
from the intersecting-angled D-strings}
\label{301}

Now let us examine the production rate for the  
massless open strings, i.e. photons,
from the intersecting D-strings. Thus, we 
should employ the critical dimension $d=26$.  
With the help of Eq. \eqref{14} and $\mathbf{Y}=0$,  
we obtain $\tilde\nu=\sqrt{2}$. 
The resultant rate is given by 
\begin{eqnarray}
\mathcal W_0 &=& \frac{V_1}{\pi V_{1+1}}
\frac{\sinh(\sqrt{2}\pi)}{\sqrt{\sqrt{2}
\pi^{-1} \alpha'}g_s^2}
\;R\left(\mathcal{E}_{(1)},\mathcal{E}_{(2)}\right) 
\nonumber\\
&\times& \sum_{m=1}^\infty \left\{
\frac{(-1)^{m+1}}{m^{3/2}} \ 
 \prod_{n=1}^\infty \left[ 
1- \exp\left(-\sqrt{2}nm
\right)\right]^{-24} \right\}, 
\end{eqnarray}
where the dependence on the electric fields 
has been collected in the function 
$R\left(\mathcal{E}_{(1)},\mathcal{E}_{(2)}\right)$,
\begin{eqnarray}
R\left(\mathcal{E}_{(1)},\mathcal{E}_{(2)}\right)
=\left(\frac{{\left(1-\mathcal{E}_{(1)}^2\right)
\left(1-\mathcal{E}_{(2)}^2\right)}}
{1-\left[\mathcal{E}_{(1)}\mathcal{E}_{(2)}
+\cosh\left(\sqrt{2}\pi\right)
\sqrt{{\left(1-\mathcal{E}_{(1)}^2\right)
\left(1-\mathcal{E}_{(2)}^2\right)}}
\;\right]^2}\right)^{1/2}.
\end{eqnarray}
In the allowed square region 
$-1 < \mathcal{E}_{(1)},\mathcal{E}_{(2)} < 1$,
this function does not possess any smooth 
maximum or minimum. Besides, 
since $\cosh\left(\sqrt{2}\pi\right)$ is a large number, 
at least one of the electric fields should be nonzero. 
Otherwise, this function becomes imaginary, which is forbidden.
In addition,
according to Eq. \eqref{9} (with the replacement 
$\tilde\nu \to \sqrt{2}$), 
such value of $\tilde\nu$ leads to
$\mathcal{E}_{(1)} \to 1$, or 
$\mathcal{E}_{(2)} \to 1$ and or both of them.
When for example $\mathcal{E}_{(2)}$ vanishes, 
for any value of the angle $\Phi$ we receive 
the following allowed values for 
$\mathcal{E}_{(1)}$, i.e.,
$\tanh\left(\sqrt{2}\pi\right) 
\leq |\mathcal{E}_{(1)}| < 1$. 

%%%%%%%%%%%%%%%%%%%%%%%%%%%%%%%%%%%
\section{Conclusions}
\label{400}

We utilized the boundary state formalism to 
compute the pair production rate of the open strings  
from the interaction of two non-intersecting
D$1$-branes at angle. 
The distance between the D-strings is arbitrary.
The D-strings have been dressed with the different $U(1)$ 
gauge potentials. Besides, the Kalb-Ramond field 
as a background was introduced. 
As a special case,  
we obtained the production rate of the 
photon pairs from the intersecting D-strings.

We observed that the production rate of the heavy 
open strings is very small, as expected. 
This is due to the fact that the required energy 
for the production of the heavy string pairs is large.
Note that 
by increasing the distance between the D-strings and 
also by enhancing of the spacetime dimension
we receive the heavy open strings.   

We observed that for acquiring the open 
string pair creation, at least one of the 
D-strings should be dressed with a nonzero 
electric field. Precisely, 
at least an electric field is obviously required 
to polarize the region between the branes.
Otherwise, the pair production cannot occur.

%%%%%%%%%%%%%%%%%%%%%%%%%%%%%%%%%%%%%%%%%%%%

\end{document}